\documentclass{sigcomm-alternate}

\usepackage{cite,amsmath,graphicx,epsfig,subfigure,times,latexsym,algorithm2e,verbatim}

\begin{document}

\title{Network Map Reduce}

\numberofauthors{1} 
\author{
\alignauthor Haoyu Song, Jun Gong, Hongfei Chen\\
      \affaddr{Huawei Technologies}\\
}

\maketitle
\begin{abstract}

Networking data analytics is increasingly used for enhanced network visibility and controllability. We draw the similarities between the Software Defined Networking (SDN) architecture and the MapReduce programming model. Inspired by the similarity, we suggest the necessary data plane innovations to make network data plane devices function as distributed mappers and optionally, reducers. A streaming network data MapReduce architecture can therefore conveniently solve a series of network monitoring and management problems. Unlike the traditional networking data analytical system, our proposed system embeds the data analytics engine directly in the network infrastructure. The affinity leads to a concise system architecture and better cost performance ratio. On top of this architecture, we propose a general MapReduce-like programming model for real-time and one-pass networking data analytics, which involves joint in-network and out-of-network computing. We show this model can address a wide range of interactive queries from various network applications. This position paper strives to make a point that the white-box trend does not necessarily lead to simple and dumb networking devices. Rather, the defining characteristics of the next generation white-box are open and programmable, so that the network devices can be made smart and versatile to support new services and applications. 
  
\end{abstract}

\section{Introduction}\label{intro}

Networking data analytics is a big data problem. The data include not only the pass through traffic, but also the states and statistics in network nodes. It has been predicted the annual global Internet traffic will pass the zettabyte mark by the end of 2016~\cite{cisco-index}. The network devices can also generate huge amount of data aggregately, if not more, from log files, databases, mirror ports, and network monitoring standards. Effective network control and management rely on good network visibility. The ability to access and understand network data is the key to gain such visibility, and in turn, to benefit network applications. For example, the traffic pattern can help to detect and predict network intrusions; the network congestion status can be used to optimize the traffic routing; the application awareness allows service providers to monetize their network offering through differential services. Numerous products such as Network Performance Management (NPM) systems and packet brokers exist. People starts to apply latest big data technologies such as machine learning to mine value from network data~\cite{ietf}.  

In addition to the sheer quantity, most network data are transient and realtime streaming data by nature. This adds another degree of challenges for applications to utilize the data, because applications may have different requirements on data and data analysis approaches. Timeliness, accuracy, and cost are all consideration factors. So far the network data analysis system is segregated from the network system itself, deeming the network only as a raw data source. The two discrete systems communicate through ad hoc protocols and interfaces. For example, sFlow~\cite{sflow} can generate digested report for sampled traffic; a packet broker uses dedicated mirror ports on network devices as the raw packet source. While this approach works, it has some drawbacks: 
\begin{itemize}
\item The ad-hoc protocol and interface fall short to provide a universal and organic network data analytics platform, which is a desire from SDN's perspective. Each analytical task may use a different protocol or interface to collect data. The two decoupled systems can only communicate through non-uniform and problem-specific APIs.
\item The value density of network data can be low. Transferring raw data from network devices not only wastes device resources (e.g., ports) and bandwidth, but also poses a heavy burden (i.e., computing and/or storage) on the analytical system that consumes the data.
\item Transferring raw data to remote analytical system also incurs long latency which impairs the capability of realtime control feedback. Imagine an interactive data analytics scenario where an application needs to quickly adjust the data capture strategy based on realtime network states.
\item Limited by the resource and bandwidth, even the raw data still have to be filtered and sampled. The inevitable information loss impairs the analytical accuracy or even renders incredible results. Limited by the data plane capability, some data required for analytics may not be available. Inferring the data from raw traffic is expensive or impossible. 
\end{itemize}

To design an efficient and organic network data analytical system, we follow two design principles:
First, to improve the analytical efficiency and reduce the processing latency, we should make the actual computing happen as close to the network data plane as possible. That is, direct in-network computing is preferred.
Second, the chief task of data plane is forwarding packets. The forwarding performance needs to be given the highest priority. The data analytics should not impede the normal data plane operation. Therefore, we should try to push the network data analytics out of the network data plane as much as possible. 

\begin{figure}[!ht]
\centering
\includegraphics[width=0.6\columnwidth]{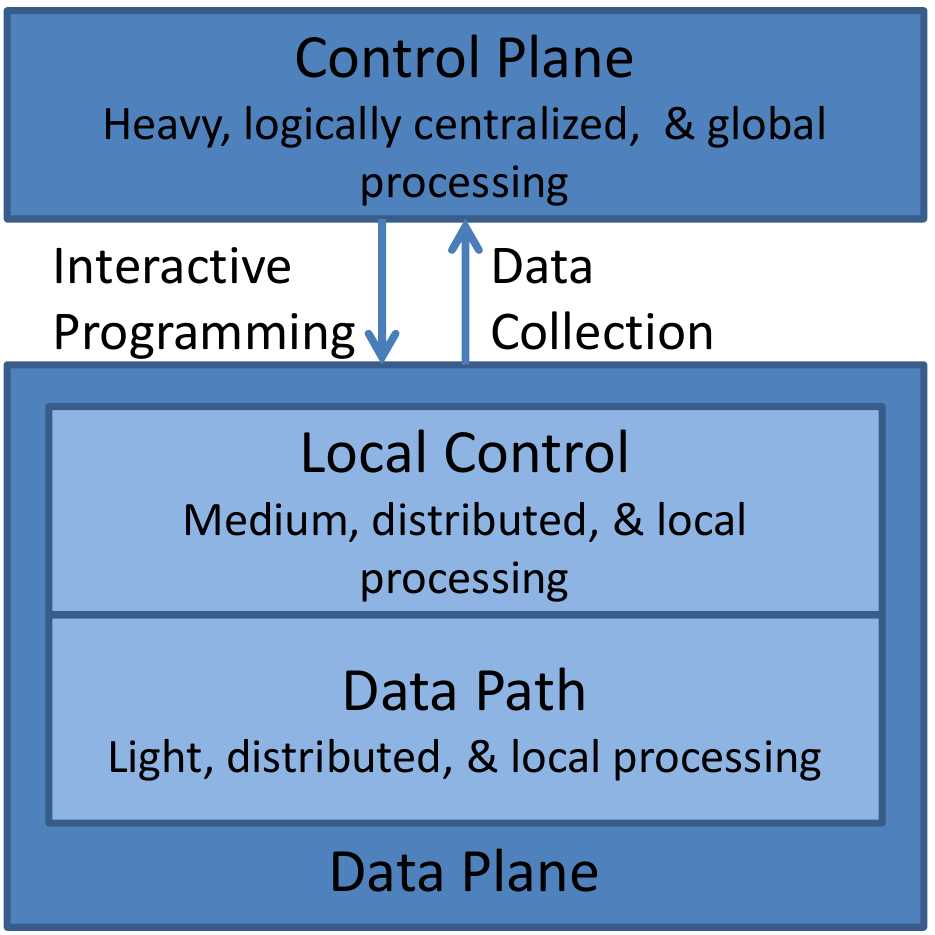}
\caption{Ideal Architecture for Network Data Analytics}
\label{fig_arch}
\end{figure}

On the surface the two principles appears to conflict with each other. We strike the balance by providing the tradeoff in the SDN architecture, as shown in Figure~\ref{fig_arch}. The analytical applications are partitioned and spread throughout all the three layers in SDN and at each layer, the tasks take full advantage of the local capabilities and avoid interfering with the other incumbent tasks. A network data analytical application involves multiple network nodes and require their cooperation to achieve the final results. The data path forwarding chips only handle local data and packets and apply light processing such as counting, filtering, sampling, and digesting. The chips then hand off the pre-processed data to the data plane local processor. The local processor is capable of doing medium level data processing to further prepare the data. The data from the distributed network nodes are aggregated at the SDN control plane for final global processing. Since the control plane is comprised of servers with high computing power, it is capable of heavy data processing and archiving.  

The question is \emph{how we should model this architecture to make it convenient for programming and configuration, and make it applicable for a wide range of network data analytical applications}.

Clearly we need a way to split an analytical application into three parts and map them into the three layers in Figure~\ref{fig_arch}. Due to the distributed nature of both the control plane and the data plane, this is also a distributed programming problem. Since the analytical applications are all runtime dynamical tasks and many of them would require modifying the data path functions and starting new computing processes in the local control processor, we need an interactive programming interface to dynamically load the data plane analytical modules into the forwarding chips and local control processors without interrupting the normal data plane processing. 

Therefore, in this paper, we draw the similarities between SDN architecture and the MapReduce programming model (Sec.~\ref{compare}) and assert that Network Map Reduce (NMR), the combined outcome of the two independent concepts, can be a powerful and efficient architecture to solve a wide range of network data analytical problems (Sec.~\ref{usecase}). This architecture is only possible with the latest data plane innovations such as dynamically programmable forwarding chips and supercharged device local control (Sec.~\ref{inno}). We lay out the system design (Sec.~\ref{design}) and preliminary implementations (Sec.~\ref{imp}). We briefly discuss the related work (Sec.~\ref{related}) and lay out the challenges and future work (Sec.~\ref{conclude}). 

With this work, we also intend to provoke a more profound discussion on the future data plane evolution. 
Currently, many believe we should push the intelligence out of the network data plane as much as possible. 
But we try to make an argument that the white-box trend is not necessary to lead to simple and dumb networking devices. Instead, the defining characteristics of the next generation white box should be open and programmable. Complicated and stateful functions can be installed in the data plane devices through open programming interface. The processing affinity is critical for high performance and low system cost. Enhanced in-network computing is essential for agile services and adds new value to network devices. It is our hope that this work can inspire more research in the direction of smart and programmable in-network computing.

\section{SDN and Map Reduce}\label{compare}

MapReduce, a programming model for massively parallel data processing and the associated implementation framework, was made popular by Google~\cite{googlemr}. It was later implemented in open source software such as Hadoop~\cite{hadoop} and Spark~\cite{spark}.

The core of MapReduce is abstracted into two primitive functions: $map(\cdot)$ and $reduce(\cdot)$, each running in a server cluster. We call a server node running the map function a Mapper, and a server node running the reduce function a Reducer. A job tracker is responsible for coordinating the task partitions and interactions between Mappers and Reducers and monitoring the task progress. Basically, each Mapper is fed a portion of the input data. The mapper tokenizes its input into $\{k_1, v_1\}$ pairs and transforms them into $\{k_2, v_2\}$ pairs. The intermediate $\{k_2, v_2\}$ pairs are shuffled to the reducers. This step ensures the pairs with the same key are located on the same reducer. Each reducer sorts and groups the list of $[v_2]$ for each $k2$ and works on the $\{k_2, [v_2]\}$ pairs to generate the list of $[v_3]$ as the final output.

The elegance of MapReduce framework is in that users only need to focus on writing the $map(\cdot)$ and $reduce(\cdot)$ functions, while the runtime distributed computing details are totally hidden. MapReduce is used to solve a wide range of big data analytics problems.

On the other hand, SDN has become the new paradigm of network architecture. A logically centralized controller is responsible for monitoring and managing the distributed network data plane. Network applications and services run on top of SDN controller and communicate with the network infrastructure through standard APIs or other configuration interfaces. 

\begin{figure}[!ht]
\centering
\includegraphics[width=\columnwidth]{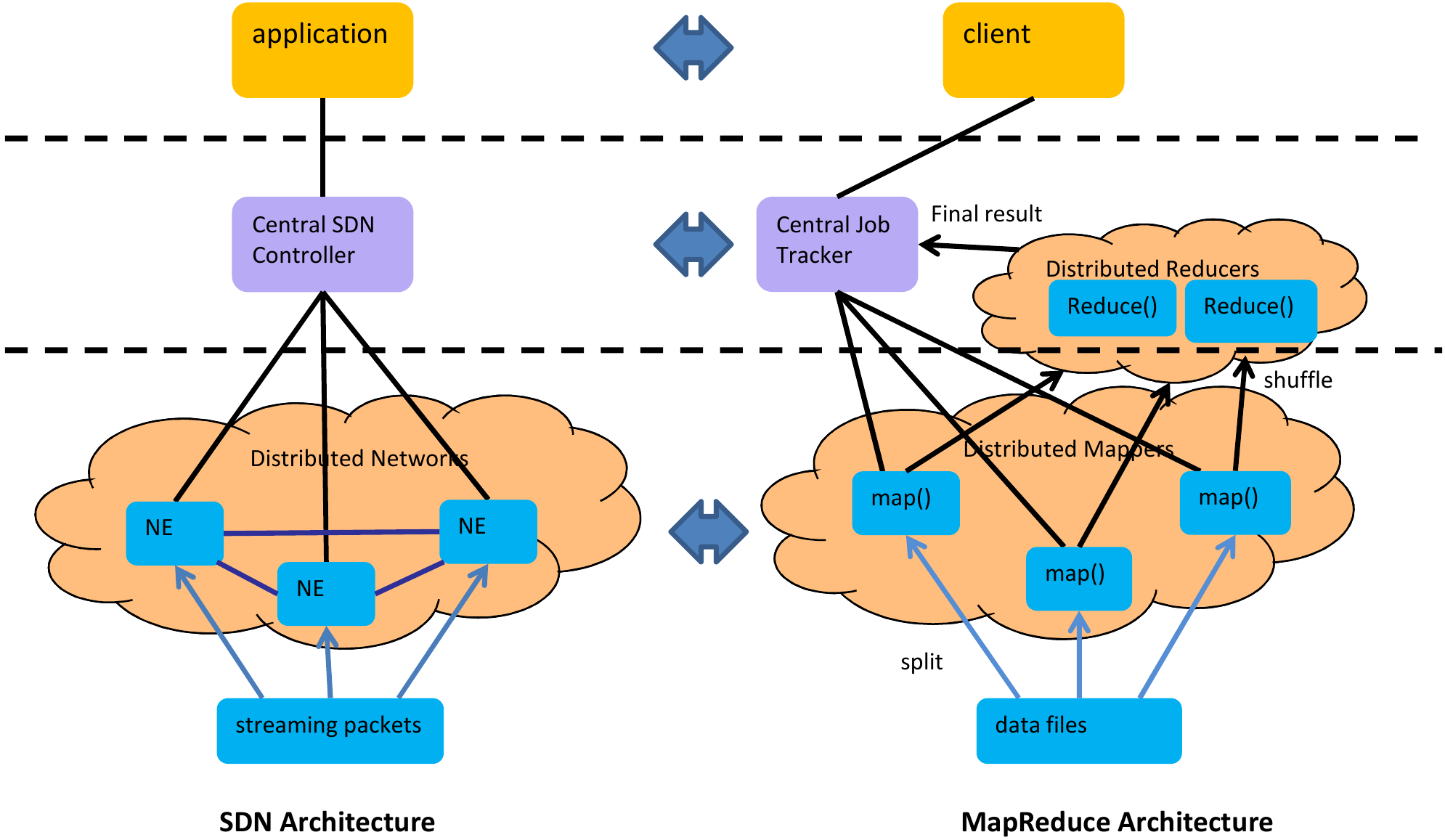}
\caption{Compare SDN and MapReduce Architectures}
\label{fig_comp}
\end{figure}

Comparing the architecture of SDN and MapReduce, as shown in Figure~\ref{fig_comp}, we can see several remarkable similarities.

Both architectures have three layers. The most important insight is that the mapper layer and the network data plane are both distributed and able to see partial input. When each mapper can work in parallel and independently with each other, each network data plane element (NE) also works independent with other NEs. An NE only processes and forwards the streaming packets coming into it. Moreover, while the SDN controller configures and oversees the entire network data plane, the MapReduce job tracker dispatches the job and monitors the overall job progress. The SDN controller has the similar role as the MapReduce job tracker. Finally, network application or service is deployed through SDN controller, and map-reduce job requests are issued from clients. Both architectures have a cross-layer control loop in which an upper layer can issue requests to and get reply from a lower layer. 

Better network visibility is considered a key SDN benefit. The visibility can be gained through comprehensive network data analytics. Since the SDN architecture bears such significant similarities with the most popular big data analytical model and framework, MapReduce, one logical question to ask is: \emph{can we embed MapReduce into SDN and make network data analytics as simple and elegant as MapReduce}? 

The answer is positive.  It is prohibitively expensive for the SDN controller to directly see and work on the entire network data, but each distributed NE node is capable of processing its own share of data, without needing to know the aggregated results. The local processing ensures the timeliness of data preparation. The preprocessed data is also significantly reduced in quantity and enhanced in quality. The SDN controller now is at a better position to gain the global network view by collecting data from NEs and conducting final analysis. Imagine we have an NMR framework, in which all we need to do is to correctly describe the network version of map and reduce functions and rely on the automatic runtime system to execute our network data analytical jobs.

Before going farther, we acknowledge that there are also some distinct differences between SDN and MapReduce.  
First, network data are streaming data which are continuously evolving cross time, but MapReduce takes static block files as data input and performs batch processing. It is possible to store the network data first and manage the data on file systems such as HDFS~\cite{hadoop}. However, network data analytics is often a continuous and incremental process and sometimes has stringent realtime requirements, so the batch processing on files is not recommended. 

Second, while an NE can be conveniently considered a mapper and the controller the job tracker, the position of reducers are not determined yet. There are basically two options: either the NEs also serve as reducers, or a standalone reducer cluster is considered as a part of the SDN controller. In the first option, an NE may need too much computing power to handle the reduce function in addition to the map function. the intermediate data shuffle also consumes extra data path bandwidth. However, as shown in Section~\ref{usecase}, usually not all the NEs are selected as mappers for an application, so the off-duty NEs can be used as reducers. Even if an NE has to run both map and reduce functions, one of the functions is usually less computing intensive and sometimes is even absent. Since the intermediate data are already highly reduced compared with the original raw data, transferring the data between mappers and reducers does not consume too much network bandwidth and impact the normal network operation. But if these are concerned, we can resort to the second option. It is possible because the SDN controller is only logically centralized and a cluster of reducers can be perfectly fit in the scope of the controller. This option consumes more control path bandwidth than the first one but it is still much better than transferring raw data to the controller as in the disaggregated analytical solutions.    
 
Third, MapReduce only produces accurate results but for many network data analytical applications, the approximate results are sometimes acceptable as a tradeoff for timely processing.

Fourth, some network data analytical applications may not exactly work in the traditional MapReduce manner (i.e., work on the $\{k, v\}$ pairs), but as long as these applications can follow the local processing plus global processing model, we can still easily partition and program the functions following the same convention.  

Finally, there are some performance issues with the conventional MapReduce implementations. For examples, some blocking operations and the reducer pull mode increase the processing latency which are not suitable for realtime analytics. These issues need to be amended for streaming data one-pass analytics. 

Nevertheless, the architecture similarity is dominant. Can the similarity leads to a better solution for network data analytical problems? We examine the popular data analytics problems in networks and confirm that NMR is indeed a viable architecture and programming model. But it does need the support from the next generation network data plane.

\section{Data Plane Innovations}\label{inno}

Many people believe the network is no more than a fabric of dumb pipes: network devices should be made as simple as possible while keeping intelligence at edge or even in terminals. The so-called ``white box'' reflects such a belief. However, the network is a complex system and always lags behind the bandwidth and delay requirements of applications. More and more network functions and services are instantiated and chained in networks. Without good network visibility, it is difficult for network operators to diagnose faults, optimize utilizations, and prevent intrusions. The network visibility is best to be gained through direct network data analytics. So in addition to forwarding traffic, the network devices need at least to be augmented with the capability to facilitate efficient data analytics. 

Since the analytical tasks are diverse, dynamic, and ad hoc, it is unlikely we can fix all of them in the network devices at design time. Rather, these tasks should be programmed into the network devices on demand and at runtime. We believe an open programming interface is needed to instill intelligence, such as in-network computing and storage, back to networks. While network devices can handle only a part of data analytical applications, some key performance indicators, such as response latency, throughput, and the cost of raw data transfer, are all spontaneously improved.

The fundamental reason hindering us from implementing an in-network analytical architecture is that the current network is rigid and 
non-programmable. In the first wave of SDN movement, the network data plane devices are still equipped with fixed-function chips and low performance local processors. A more flexible and powerful data plane is needed and two key enabling technologies are discussed as follows.

\subsection{Programmable Forwarding Chips}

Luckily, flexible and programmable data plane is being developed and will be steadily available soon~\cite{pof-1, p4}. The core component on network data path is forwarding chip. Chips like CPU, NPU, and FPGA are essentially programmable. Programmable ASIC is also proved feasible and being built by some vendors~\cite{stanfordTI}. The logical next step is to open up the chip programming interface to network users. High level language such as P4 [10] is developed to incarnate the network devices into different functioning boxes depending on the program installed.

However, such programming process is considered static, because even a
minor modification would require the entire source code to be recompiled and the application reinstalled. Dynamic network data analytics asks for prompt reaction. The applications should not interfere the normal traffic forwarding in any case. 
The static process not only incurs long deployment latency but also temporarily break the normal forwarding operation during the data path reconfiguration. Hence, we believe the forwarding chips should be further re-architected to support runtime interactive programming. 

We have developed the Dynamic Network Probes (DNP) as passive network data processing and collecting method mechanism~\cite{dnp-a}. DNP is based on the POF programming model~\cite{pof-2} and takes advantage of the POF interface's interactive programming capability. A DNP can realize counters, meters, filters, and even Finite State Machines (FSM) for event monitor. DNPs can be dynamically deployed at various data path locations including ports, queues, buffers, and flow tables. We will show DNP is exactly what is needed at network data path level to build an NMR system.
 
DNP is not necessarily the only way to support NMR. Open source projects such as IOVisor~\cite{iovisor} and fd.io~\cite{fdio}, working on virtual network devices, provide other means to dynamically program and load probes or other new function extensions on data path at runtime.

\subsection{Server-Grade Local Control}

Programming the forwarding chip is not enough to realize a full NMR system, because the chips can only finish the preliminary data preparation. We still need the local control processor's cooperation to further process the data.  

The local control processor is the brain of network devices. In conventional network devices, it handles limited tasks which are not compute intensive. For example, the processor in a router is mainly used to process routing protocols and update forwarding tables, so the mediocre CPU and memory are sufficient. The data transfer between the control processor and the data path is also limited. 
However, for the data analytical architecture shown in Figure~\ref{fig_arch}, the bandwidth requirements between the local control processor and the data path, and between the data plane and the control plane are both significantly increased. The local processor directly participates in data processing which is much more compute intensive. The local processor even needs persistent storage to cache data for future reference. All these require us to boost the performance of the local processor by providing server-grade local control. The local control and the forwarding chips are interfaced by high bandwidth Ethernet ports.       

The computing cost keeps becoming cheaper and the potential benefits can justify the extra cost of the supercharged local processor. Pluribus pioneered the idea of server-switch which provides a high performance local control plane~\cite{sswitch}. Faster CPU, larger memory, and local storage make a switch as powerful as a server. Instead of scaling up the local processor, we can also take the scale-out approach. For example, in data centers, we can always use the spare servers or VMs in the same rack as the local control processor of the ToR switch. Similarly, NoviFlow developed the scale-out routers which add servers in router chassis to enhance the local data plane~\cite{noviflow}. These examples share the same scale-out spirit and can be applied depending on the device location. 

The local processor's programmability is a given. Applications can be easily deployed as virtual network functions (VNFs). Each VNF is essentially a microservice so we can use technologies such as container to quickly install and update VNFs and isolate them. 

\section{Design}\label{design}

The two network data plane techniques, when combined together, provide us the foundation to build an integrated NMR system. The new generation of network data plane is naturally a distributed mapper cluster. It can also be optionally overloaded as a reducer cluster. The network control plane assumes the responsibility of the job tracker. It also communicates with applications, generates the map and reduce functions, compiles and installs the functions into different components.

\begin{figure}[!ht]
\centering
\includegraphics[width=0.4\columnwidth]{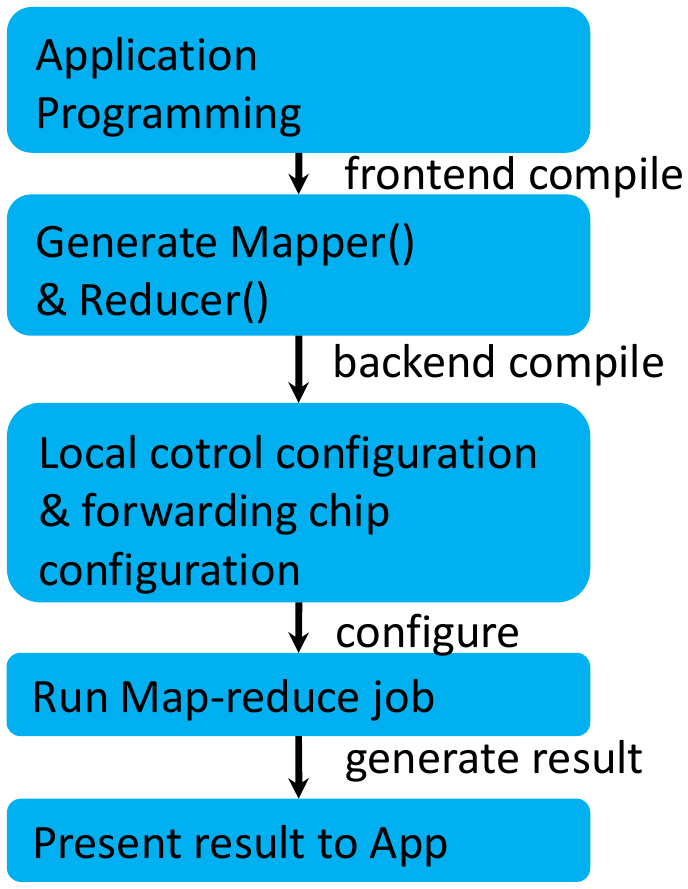}
\caption{The Work Flow of Network Map Reduce}
\label{fig_flow}
\end{figure}

The overall work flow of NMR is shown in Figure~\ref{fig_flow}. A data analytical application is first programed and compiled into a map function and a reduce function. These functions are target agnostic. A target specific back-end compiler will further compile the map function into two parts: one part is used to configure the forwarding chip and the other part is used to configure the device local control processor. Next, these configurations are downloaded to the network data plane. If the data plane is also used as the reducer cluster, the reduce function is also downloaded to the network data plane. Otherwise, the dedicated reducer server cluster is configured. After these steps, the actual map-reduce job can be started to generate results. When the SDN controller receives the results, it will pass the results to the application. 

As we will show later, usually not every data plane device needs to be involved in an application. The map function is also responsible for picking the subset of network devices to participate in the computing. The reduce function can be deployed to the network devices which are not configured as mappers. The reduce function only runs in the local processors of these network devices.    

Since network data plane is essentially a steaming system, the NMR must adapt to the streaming data processing mode. Some applications only request one-shot result but many applications request continuous data analysis. The mapper only conducts stateless data filtering and transformation on the continuous data stream. Therefore, the output of mapper can be the intermediate streaming data too. Instead of waiting for the reducer to pull the data, the data are pushed to reducer as soon as they are available. The reducer runs sliding window or jump window on the intermediate streaming data to generate the final results. Both mappers and reducers have storage capability so they can be configured to cache some data in order to handle queries which require historical data. 

The map and reduce functions running in the local processor are configured as microservices, so multiple NMR jobs can run in parallel to respond queries from multiple applications. These functions may subscribe to the same data so the data need to be properly labeled for recipients.       

Depending on applications, the map and reduce functions could be very simple or even absent. Nevertheless, the model is helpful to simplify the job and make the computing happen close to the data source.

\section{Applications}\label{usecase}

The network data analytics finishes the ``visibility'' part of the SDN control loop. NMR is a convenient and efficient model and architecture. As long as we can model an application as distributed data plane preprocessing plus logically centralized data aggregation, we can use NMR to solve it. The network data analytical applications can be categorized into four groups: QoS, security, customer care, and optimization. QoS is about the network performance measurement and application analysis. Security is about network intrusion detections using techniques such as deep packet inspection and stateful flow monitoring. Customer care is about customer behavior analysis and trouble shooting. Optimization covers applications such as network load monitoring, traffic matrix measurement, and elephant flow identification, which are used to provide inputs to the traffic engineering algorithms.   

Here we provides a few simple examples to show how NMR works. For these applications, the map and reduce functions are all dynamically loaded to the network data plane and control plane.  

\subsection{DDoS Detection}

In a data center we want to find the servers under possible DDoS attack with a suspiciously large number of connections. First we picks all the portal switches in the network as mappers and choose a few free switches as reducers.

The map function has two parts. It configures the forwarding chips to filter all unique flows to the pool of servers under monitoring. It configures the local processor to calculate $\{k, v\}$ pairs for a predefined time window and send the data to reducers, in which $k$ is the server id and $v$ is the number of unique flows targeting to that server. 

The reduce function is simple: it calculates the global $\{k, sum(v)\}$ pairs by summing up the flow count for each server in each time window. If for any server, the number of unique flows exceeds a predefined threshold, an alarm is triggered and sent to the application.
We can also measure the load of target servers by slightly modifying this application. 

\subsection{Traffic Matrix Measurement}

In a WAN we want to measure the traffic matrix in a predefined time window. The traffic matrix includes all the source and destination router pairs (i.e., the key) and the bytes and packets flowing through each pair of routers in a time window (i.e., the value). In this case we pick all the edge routers $r_1 ... r_n$ in the network as mappers. We can choose a few other free routers as reducers.

The map function has two parts. The forwarding chip in $r_i$ needs to label each ingress packet with the router ID $r_i$ and keeps the byte and packet counters for all the egress packets from $r_j$. The local process periodically reads and resets the counters, and pushes the ${k, v}$ pairs to the reducers. 

The reduce function summarizes all the inputs from mappers, generates the traffic matrix, and reports to the application. 

In addition to the traffic count, the path latency between each pair of routers can also be tracked. We just need to ask the forwarding chip to timestamp the ingress packets and calculate the latency of the egress packets, and ask the local processor to maintain the minimum and maximum latencies for each router pair. 
                            
\subsection{Network Congestion Monitor}

Network congestion is reflected by packet drops at routers. While it is easy to get the packet drop count at each router, it is difficult to gain insights on the victims, hot spots, and lossy paths. We can deploy an NMR application to acquire such information. All routes in the network are chosen as mappers. Some routers or the dedicated server cluster are chosen as reducers.

The forwarding chip in each mapper will notify the local processor the flow signature and the port for each dropped packet. The local processor generates the drop statistics for flows and ports in continuous time windows. The data are pushed to the reducers. The reducers aggregate the input data to generate the report on the top victims, hot spots, and the most lossy paths. 

\subsection{Elephant Flow Detection}

We want to track network-wide top-$n$ flows. Various algorithms has been developed at each NE to detect local elephant flows~\cite{Estan, elephant, flowradar}. We choose a set of mappers to run one of the algorithms as map function, and choose one or more reducers to aggregate the $\{k, v\}$ pairs from the mappers, where $k$ is the flow id and $v$ is the flow statistics. Note that this method can also catch the sneaking elephant flows which take multi-path forwarding strategy. In a mapper, the fast path data structure is programmed in the forwarding chips and the slow path data processing in programmed in the local processor. 

In some cases, the local resource in an NE is not sufficient to monitor the entire flow space. We can partition the flow space and configure each mapper to track only a subset of flows, given the assumption that each mapper can see all the flows. The reducers will rank the inputs from mappers to pick the global top-$n$ flows.

\section{Implementation}\label{imp}

In our preliminary proof of concept implementation of NMR, we handcraft the code for the map and reduce functions. The map function contains two parts: one part is the modification to the forwarding chips and the other part is the code running in the local processor of the network device.  

\begin{figure}[!ht]
\centering
\includegraphics[width=0.9\columnwidth]{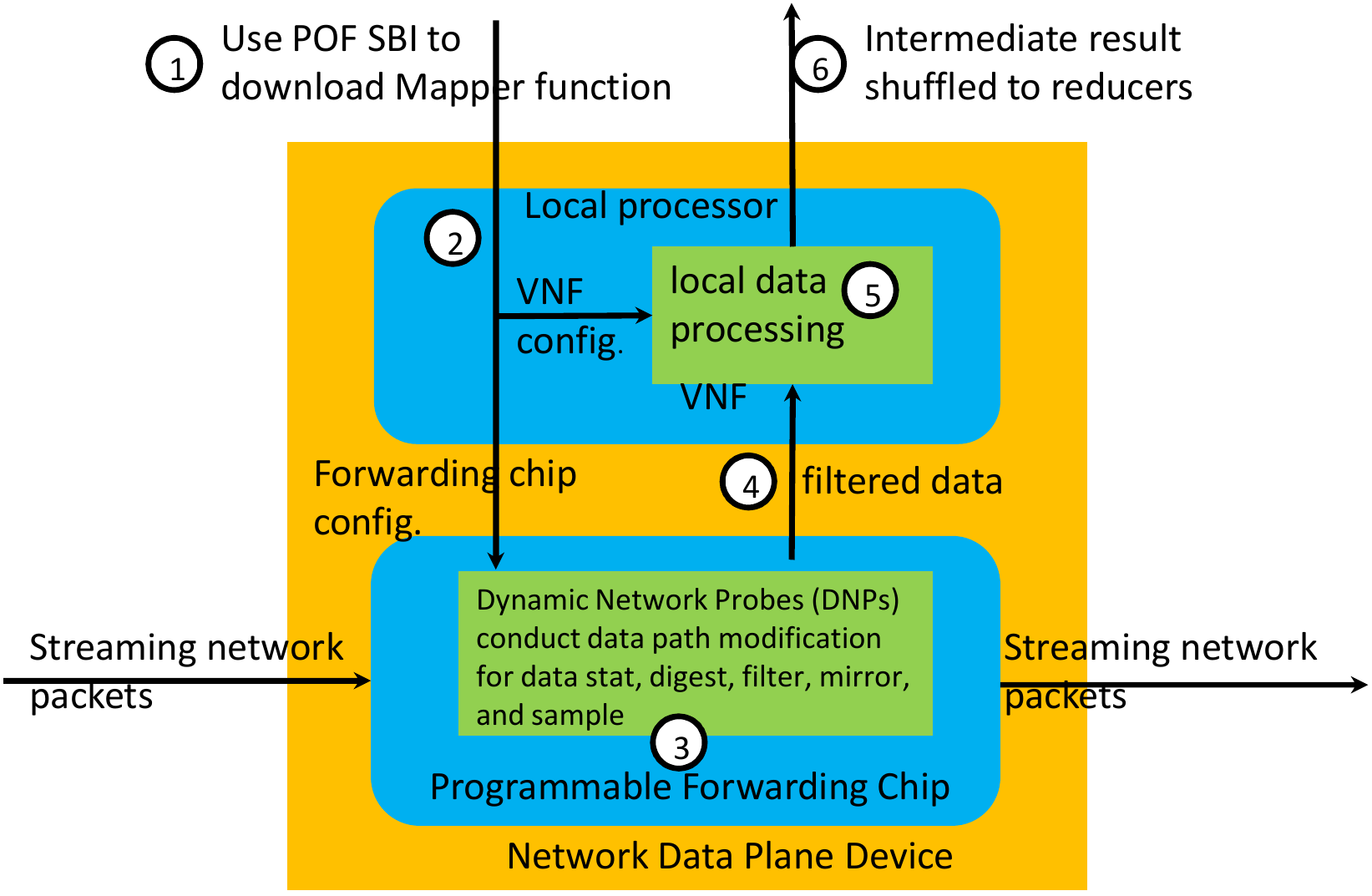}
\caption{Implement Mapper in Network Device}
\label{fig_map}
\end{figure}

The core component in NMR is the mapper, implemented in network data plane devices. Our prototype is a router platform in which the forwarding chips on line cards are network processors (NPU) and the local processor is a PowerPC-based subsystem. The POF runtime interface~\cite{pof-2} is used to download the map function to the device. The part of map function that configures the forwarding chips is compiled into DNPs~\cite{dnp-a} and the part of map function that configures the local processor is initialized as a process. Our prototype uses the SDN controller as the reducer and the intermediate data from the mappers are pushed to the reducer through the POF runtime interface. The detailed work flow is shown in Figure~\ref{fig_map}.  

We are working on a formal programming language and the compiler which can automatically partition and load the functions, track the processing progress, and collect the results. The NMR system should be easily hooked up with other applications to complete the SDN control loop. 

\section{Related Work}\label{related}

Streaming network data analytics has long been studied with the high scalability and low latency as the main optimization goals\cite{model-stream}. MapReduce is considered a proper model for streaming network data analytics~\cite{Lee-survey, Lee-hadoop}, but so far no attempts has been made to embed it in network (or at least partially) due to the lack of data plane programmability and processing power. However, the findings and approaches revealed by the previous research (e.g., Spark Streaming~\cite{sparkstream}) can still be applied in the NMR system.
   
Stream MapReduce combines the MapReduce programming model and the event stream processing to achieve scalability and low latency~\cite{smr}. The mappers are stateless filters and the windowed reducers are stateful operators. The evaluations show multi-fold response time reduction and throughput increase. NMR can add extra cost and latency benefits on top of this.

Muppet is an implementation of MapUpdate, a MapReduce-like framework dedicated for fast data~\cite{muppet}. In this framework, the reduce function is replace by an update function. The $\{k, v\}$ pairs are augmented with the stream id and time-stamp. The update function use the notion of slate to summarize all the past events and update the current state from the mapper inputs.

MapReduce Online~\cite{mronline} modifies the MapReduce architecture to allow data to be pipelined between operators so the resulting framework is suitable for continuous queries on stream data. S4~\cite{s4}, inspired by MapReduce model, is a general purpose platform for continuous and unbound streaming data processing. In~\cite{opa}, MapReduce model is also modified to support incremental one-pass analytics so the online aggregation and stream query can be conducted near real-time. These ideas can be absorbed by NMR.

Data analytics in networks does not necessarily use the MapReduce model. SQL-like queries, while adapting to streaming data, is also possible~\cite{model-stream, stream-query}. Some applications run machine learning algorithms on various network data, trying to figure out automate management rules~\cite{net-automation, ietf}. However, we believe NMR can still apply because it is always more efficient to ask the data plane devices to pre-process and ``compress'' the data so the control plane only needs to work on the structured and relevant data. 

Before NMR, network devices are only considered separated raw data source for network data analytics~\cite{sflow, planck, flexam, everflow, netsight}. Some techniques use in-network computing to process or pre-process network data~\cite{flowradar, minions, osketch, ostate}. These techniques can solve some specific data analytical problems but do not provide a uniform programming model like NMR. The data plane algorithms can be implemented in the map or reduce functions.  

SDN is also useful for big data applications in that the runtime network configuration can closely interact with the big data applications for joint performance optimization. An integrated control architecture for SDN controller and MapReduce scheduler is discussed in~\cite{sdnbigdata}. 

\section{Conclusion}\label{conclude}

We show the SDN architecture can mimic the MapReduce model and propose NMR, an integrated network data analytics platform which embeds the MapReduce programming model into SDN, while taking advantages of the latest network data plane innovations. The scalability of the solution is naturally achieved by the scale of the network data plane and the clean job partitioning. We model many of the popular network management and monitoring applications into MapReduce-like jobs and execute them in NMR at low cost and low latency. We believe more interesting applications can be conceived under this framework. 

NMR is just an example that the in-network programmable computing can bring significant values to network service providers and users. The model fragments applications into multiple interdependent pieces which are then distributed into different execution environments (e.g., CPUs and forwarding chips). However, the high level application is not aware of such details. The job partition, allocation, and management are performed automatically. This relies on not only a good model but also a strong compiling system and a versatile runtime system. 

We acknowledge that we just scratch the surface of the problem space by providing a simple model and a framework. Although NMR is an enabling technology with obvious advantages, we still need to quantify the performance benefit of NMR over the conventional discrete soluiton. There are still missing pieces in our implementation which deserve future research. We hope this work to provoke more research along this line of thought and more use cases to be developed.   

\bibliographystyle{abbrv}
\bibliography{sigproc}  

\end{document}